\documentclass[a4paper,UKenglish]{lipics}
\usepackage{amsmath,amssymb,caption,keyval,subfig,graphicx,wrapfig,appendix}

\title{Ancilla Driven Quantum Computation with arbitrary entangling strength}
\author{Kerem Halil Shah}
\author{Daniel K. L. Oi}
\affil{SUPA Department Of Physics, University of Strathclyde\\
107 Rottenrow, Glasgow G4 0NG, UK\\
k.halil-shah@strath.ac.uk, daniel.oi@strath.ac.uk }
\authorrunning{K. Halil Shah and D. K. L. Oi}


\begin{document}

\maketitle
\begin{abstract}
  We extend the model of Ancilla Driven Quantum Computation (ADQC) by
  considering gates with arbitrary entangling power. By giving up
  stepwise determinism, universal QC can still be achieved through a
  variable length sequence of single qubit gates and probabilistic
  ``repeat-until-success'' entangling operations. This opens up a new
  range of possible physical implementations as well as shedding light
  on the sets of resources sufficient for universal QC.

\end{abstract}

\section{Introduction}

A fundamental question is ``what set of resources are required for
universal quantum computation?''. Many models have been proposed
ranging from the conventional gate-based~\cite{GBQC},
measurement-based~\cite{RaussendorfBriegel},
adiabatic~\cite{QuantumComputationByAdiabaticEvolution}, and
topological~\cite{KitaevAnyons} quantum computation. These models
utilise different sets of resources and are suited to various physical
implementations. By finding new models with different sets of
sufficient resources, this may open up new ways of implementing
quantum computation as well as enrich our understanding of quantum
computation itself.

A hybrid model combining aspects of the gate-based and measurement-based approaches was introduced as Ancilla Driven Quantum Computation~\cite{ADQC}. The key feature of ADQC is the restriction of resources to a single unitary interaction, and direct access (initialisation and measurement) only to an ancilla qubit that can be coupled sequentially to system (register) qubits. ADQC uses entanglement between the ancilla and register, and the kickback induced by measurement on the ancilla to drive unitary evolution of the register. By coupling the ancilla to various register qubits and choosing different measurements on the ancilla, universal QC can be achieved.

There is often a trade off between easy access and manipulation or long coherence times. ADQC lends itself to physical systems where register qubits with long decoherence times are difficult to manipulate while relatively short-lived ancillary systems are more easily controlled and can be quickly initialised and measured. There has been much work that looks at dealing with such properties with the use of ancillas for specific physical implementations. Work on optical clocks using aluminium ions have employed magnesium or beryllium ions as an ancillary system to account for a lack of optical accessibility with aluminium ions~\cite{AluminiumOpticalClock}. Ohshima~\cite{Ohshima2000} considered maintaining low decoherence of quantum dots by only activating access through an ancilla qubit in the same cell. It has been proposed that isolated, stable NV centre nuclear spins be used as qubits manipulated by neighbouring electron spins~\cite{QuantumRegisterNVCentre,UlmNVCentreAncilla} and Bermudez \emph{et al} developed an proposal for nuclear spin interactions mediated by electron spins that effects an Ising type interaction~\cite{UlmNVCentreAncilla}. Ion trap+photon systems such as in~\cite{BlinovEtAl04} and solid state systems with ballistic electrons have been considered for a class of systems that involve generating quantum gates through scattering between a static and flying qubit~\cite{FlyingQubits}. In such cases we may have no interaction-time tuning~\cite{FlyingQubits} or restricted range of dynamic modulation which is particularly relevant for the work of this paper.

These proposals often look at one system or particular parameters with a focus on gate based QC. The proposals of Anders \emph{et al}~\cite{ADQC} and Kashefi \emph{et al}~\cite{TwistedGraphStatesADQC} from which we will develop our proposal considers a scheme without reference to a specific physical system. A general description is provided, of the minimal resources require to achieve universality, expressed as a finite resource set that can be compared to Measurement Based Quantum Computation (MBQC) and the resources to form cluster states but aimed towards a Gate Based Quantum Computation (GBQC) style circuit.

In the original ADQC scheme, the computation proceeds step-wise deterministically up to Pauli corrections on the final output, very much in the spirit of MBQC~\cite{RaussendorfBriegel,BrownBriegelOWC}. This requires that the entangling unitary between register and ancilla be of either of two specific gates, $H\otimes H.CZ$ or SWAP$.CZ$. No other entangling gate would allow the measurement-induced kickback to be unitary, and permit the different computation branches corresponding to the possible measurement results to be reunited via Pauli corrections~\cite{ADQC}.

Here, we show that by relaxing the requirement for stepwise determinism a much larger set of entangling gates can achieve universal QC which may open up a range of physical implementations with fixed arbitrary coupling strengths and interaction times. Investigation of to what extent the required entangling properties can be relaxed has been performed for MBQC~\cite{NovelSchemesForMBQC,MBQCBeyondOW}. Unlike that, we do not use many-body physics techniques but expand the resource set using GBQC concepts.

 Pauli corrections and the corresponding direct access of the register will not be required. This may also reduce the control requirements for characterisation as described in~\cite{OiSchirmer2012}. This is achieved by gate approximation, repeat-until-success strategies, and multi-step measurement. Single qubit unitary gates can still be implemented deterministically while two qubit gates and measurement and state initialisation will be probabilistic and require the development of probabilistic protocols of which we provide examples.

\section{Overview of ADQC}
We will review how ADQC implements single qubit unitary gates and which conditions on the available resources are necessary for the method to be step-wise deterministic. We highlight the resource requirements that our proposal will extend; a full description can be found in~\cite{ADQC,TwistedGraphStatesADQC,ADQCWithTwisted}.

The evolution of a qubit in a quantum register can be driven by preparing an ancilla qubit system in a state $|a\rangle$, coupling that ancilla with an entangling interaction, $E_{AR}$  and then measuring the ancilla in some basis $\{|m_+\rangle, |m_-\rangle\}$. After measurement, the evolution of the register qubit can be described by a Kraus operator $K_{\pm}=\langle m_{\pm}|E|a\rangle$~\cite{NielsenChuang, KeylWernerCM, StinespringCSA}.

 The interaction $E_{AR}$ can generally be composed of a product of gates that act locally on the individual systems and a gate that produces some entanglement between the systems.  For classification of two qubit gates, we turn to the canonical decomposition~\cite{KrausCirac01,GeometricTheoryOfNonLocal,Rezakhani}: A unitary that acts on a system of a pair of qubits A and R, $E_{AR}$, can be expressed as

\begin{equation}
E_{AR}= (V_A \otimes V_R)\Delta(\alpha_x,\alpha_y,\alpha_z)(U_A\otimes U_R)
\end{equation}
where  $U_A,U_R,V_A,V_R$ are unitary gates that act only on the local systems of qubits 1 and 2 and $\Delta(\alpha_x,\alpha_y,\alpha_z)=\text{exp}(-i (\alpha_x \sigma_x\otimes \sigma_x+\alpha_y \sigma_y\otimes \sigma_y+\alpha_z \sigma_z\otimes \sigma_z))$.

If the final action on the register is to be unitary then the Kraus operator must be proportional to unitary such that $K_{\pm}K_{\pm}^{\dagger}=p_{\pm}\mathbb{I}$ where $p_{\pm}$ is the probability of the measurement. This does not depend on the local unitary gates acting on the register.  The $(\alpha_x,\alpha_y,\alpha_z)$ parameters are the only parameters unique to $E_{AR}$ that determine whether the measurement back-action will be unitary. The preparation of the state of the ancilla system can be adapted for any unitary that immediate follows it; $U_A$ is effectively under full control and can be represented entirely by ancilla state preparation. Similarly choice of measurement basis equates to freedom of choice of the unitary after interaction, $V_A$. Only the $\Delta(\alpha_x,\alpha_y,\alpha_z)$ component is of interest when considering the entangling capabilities of the interaction~\cite{KrausCirac01,GeometricTheoryOfNonLocal}.

$K_{\pm}$ are generated probabilistically and are not equivalent. In order to be deterministic, a key idea from MBQC is utilised: if $PK_-=K_+$ where $P$ is a Pauli operator correction, this correction can then be commuted through several applications of the Kraus operator and local unitary gates. Different computation branches associated with the measurement outcomes can be reunited by a Pauli correction.
\begin{equation*}
(V_R P K_- U_R).(V_R P K_- U_R).(V_R P K_- U_R) = P^{\prime}P^{\prime\prime}P^{\prime\prime\prime}(V_R K_- U_R).(V_R K_- U_R).(V_R K_- U_R)
\end{equation*}

To have the capability to enact any arbitrary unitary gate on the register, a sequence of $V_R K_- U_R$ should be universal for single qubit unitary gates.

\subsection{Control-Z Hadamard example}

\begin{center}
\includegraphics[height=3cm,width=8.5cm]{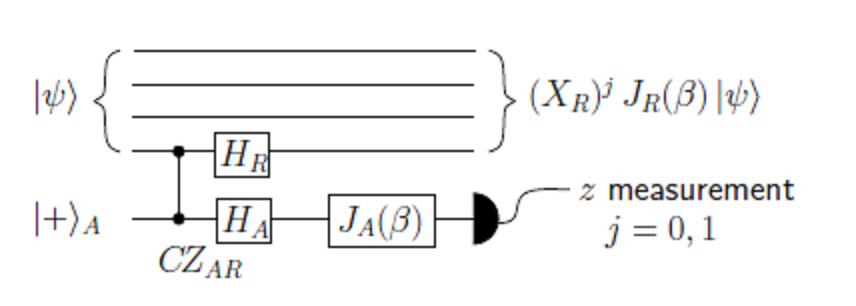}
\captionof{figure}{Anders \emph{et al.} depiction of ancilla-driven implementation of a single qubit rotation~\cite{ADQC}. The ancilla and register qubits are coupled with $CZ$ and the local unitary gates are chosen such that the interaction remains symmetrical with respect to ancilla-register exchange. A rotation $X^j J(\beta)$ is enacted on the register a result $J(\beta)$ is enacted on the ancilla which is then measured in the z basis with a result $j$=0,1.}
\label{fig:CZHadamardExample}
\end{center}

An example can be performed with $E_{AR}=(H_A\otimes H_R). CZ_{AR}$. The ancilla is prepared in the $|+\rangle$ state, couples with the register qubit with $E_{AR}$ and then undergoes a unitary $J(\beta)=H R_{\hat{z}}(\beta)$ before being measured in the computational basis. The action on the register qubit for a measurement result $|j\rangle$ is $X_R^j J_R(\beta)$. The difference between the two resulting unitary operations can be corrected by $X_R$. Any arbitrary single qubit unitary can be decomposed into four rotations $e^{i\alpha}J(0)J(\beta)J(\gamma)J(\delta)$ so to implement any arbitrary single qubit unitary up to a global phase, four ancilla interaction-measurements are performed changing the parameter of the local unitary on the ancilla to be the Euler angles $\delta$, $\gamma$, $\beta$ and then 0,
\begin{equation}
X^i J(0) X^j J(\beta)X^k J(\gamma)X^l J(\delta).
\end{equation}
 As in MBQC, the Pauli corrections can be commuted through each application of $J(\beta)$ if we make adaptations to the local unitary applied to the ancilla measurement basis~\cite{RaussendorfBriegel, BrownBriegelOWC, TwistedGraphStatesADQC}.

\begin{equation}
X^i J(0) X^j J(\beta)X^k J(\gamma)X^l J(\delta)= X^i Z^j X^k Z^l J(0)J((-1)^k \beta)J((-1)^l \gamma)J(\delta)
\end{equation}

To extend to larger computations all corrections on register qubits are required to interchange with future entangling operations $\Delta(\alpha_x,\alpha_y,\alpha_z)$ so that they remain local corrections on the register and on the ancilla preparation or measurement basis choice. This allows all the Pauli corrections to accumulate at the final step. This provides the condition that the entangling operation $\Delta(\alpha_x,\alpha_y,\alpha_z)$ tensor commutes with the corrections~\cite{ADQC}; as a result only two classes of coupling are universal.

\subsection{Conditions on the class of interactions}

In order to allow achronical Pauli operator corrections ensuring general stepwise deterministic single qubit gates exactly, we must have interactions that belong to the classes of interactions that are locally equivalent to $CZ$ or $CZ+SWAP$ gates. Expressed in terms of $\Delta(\alpha_x,\alpha_y,\alpha_z)$, these classes are $\Delta(\frac{\pi}{4},0,0)$ and $\Delta(\frac{\pi}{4},\frac{\pi}{4},0)$. There exists broader classes of interactions which fulfil the weaker condition that the Kraus operators acting on the register qubit are proportional to unitary.  These are the classes for which there is at least one $\alpha_i$ parameter equal to zero-  $\Delta(\alpha,0,0)$ and $\Delta(\alpha_1,\alpha_2,0)$ up to symmetries. It is symmetrical up to local unitary gate corrections with respect to $\pm$ exchanges, $\frac{\pi}{2}$ shifts and reflections in $\frac{\pi}{4}$ acting on the parameters $(\alpha_x,\alpha_y,\alpha_z)$~\cite{KrausCirac01,Rezakhani}. The values are also symmetrical with respect to permutations~\cite{Rezakhani}. This allows us to consider only cases where $|\alpha_j|<\frac{\pi}{4}$ and to classify an interaction by the number of non-zero parameters, e.g. $e^{-i(\alpha \sigma_z\otimes \sigma_z)}$ is equivalent, up to local unitary corrections to any case where $(\alpha_i=0,\alpha_j=0,\alpha_k\neq 0 )$.
We will demonstrate a way in which we can eschew a stepwise construction of unitary gates, thus not requiring Pauli corrections and allowing the broader class  $\Delta(\alpha,0,0)$, locally equivalent to the Control-unitary set of gates, for ancilla driven quantum computation with arbitrary interactions. As a cost, we will not be performing exact unitary gates but efficiently generated approximation.

 Some characteristics are general for any member of the class so we will use $e^{-i\alpha \sigma_z\otimes \sigma_z}$ to represent them. Other effects are dependent on specific interactions with descriptions of the local unitary gates. In the next section we will explain our choice of interaction in those cases.

\section{Single qubit gates using $(H\otimes H).C\text{-}T$}

An interaction $\Delta(\alpha,0,0)$ where $0<\alpha<\frac{\pi}{4}$ has a restricted set of conditions on the choice of ancilla preparation state $|a\rangle$ and measurement basis $\{|m_j\rangle\}$ that allow for a measurement induced back action represented by the Kraus operator $\langle m_j|\Delta|a\rangle$ to be unitary. The resulting set of unitary gates that can be implemented deterministically is smaller than for $\Delta(\frac{\pi}{4},0,0)$. Using $\Delta(\alpha,0,0)$, $\alpha<\frac{\pi}{4}$, only two gates can be implemented deterministically. If we consider just using the non-local part of the Cartan decomposition $e^{-i\alpha\sigma_z\otimes \sigma_z}$, the only two possible unitary gates that can be enacted independently of the random measurement result are achieve by preparing the ancilla in the computational basis. A $|0\rangle, |1\rangle$ ancilla input always corresponds to a $R_{\hat{z}}(2\alpha),R_{\hat{z}}(-2\alpha)$ unitary gate on the register qubit respectively.

However, with local unitary gate corrections any member of the class $\Delta(\alpha,0,0)$ is equivalent to a Control-Unitary gate. Consider this in conjunction with a fixed local unitary post-action, $U_b$, on the register. $U_{b,R}.C\text{-}U_a$ can implement a two gate finite set $\{U_0,U_1\}$ where $U_0=U_b$, $U_1=U_aU_b$. If the Lie algebra closure of $U_0$ and $U_1$ covers SU(2) then these two gates form a universal set for single qubit unitary gates~\cite{OnUniversalAndFaultTolerant}.

\begin{center}
\includegraphics[height=2.9cm,width=7.5cm]{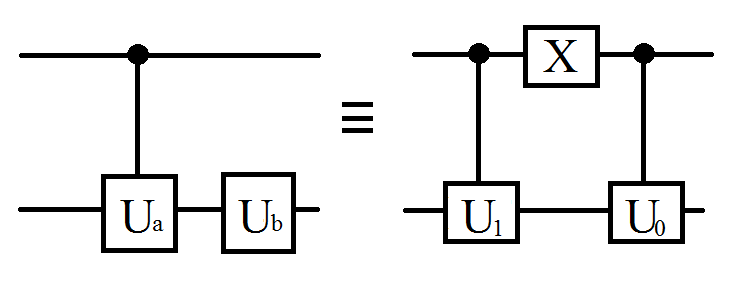}
\captionof{figure}{A generalised description of the action by which members of the interaction class $\Delta(0<\alpha<\frac{\pi}{4},0,0)$ can generate two unitary gates whose commutation rules do not form a closed set. $U_0=U_b$ and $U_1=U_bU_a$}
\label{fig:GenerallCircuit}
\end{center}
To demonstrate our proposal we choose the fundamental unitary for ancilla driven quantum computation $E=(H\otimes H). C\text{-}R_{\hat{z}}\left(\frac{\pi}{4}\right)$ (which corresponds to $\alpha=\frac{\pi}{16}$). This specific gate is chosen for two reasons; 1) It is directly comparable with the gate $E_{AR}=(H\otimes H). CZ$ from~\cite{ADQC} but with a smaller rotation angle parameter of the Control-unitary, 2) It will generate $\{T,HT\}$ which generates the same group as the finite set $\{H,T\}$ which is well studied and proven to be universal~\cite{OnUniversalAndFaultTolerant} and of interest for its applications in fault tolerant quantum computation~\cite{OnUniversalAndFaultTolerant,AliferisGottesmanPreskill}. The method can then be generalised to arbitrary coupling strengths and local unitary gate products. With this method, the circuit is programmed by the ancilla state preparation and requires no further manipulation nor measurement of the ancilla. This may have benefits for particular physical systems depending on the lifetime and robustness of the ancilla.

With this choice of interaction for producing single qubit gates, it is necessary to demonstrate the ability to perform measurements, initialise the register qubit into a specific state and enact a two qubit entangling gate in order to achieve universal quantum computation. We will address these issues in later sections.

\subsection{Two parameter interactions and stochastic ADQC}

The above method can only be employed with ``one-parameter'' interactions ---the class $\Delta(\alpha,0,0)$. There does not exist an ancilla preparation basis for which the resulting action is deterministic in the case of the $\Delta(\alpha_x,\alpha_y,0)$- ``two-parameter'' class. Instead there must be a measurement of the ancilla which results in an action that is dependent on the measurement result.

A two-parameter class interaction can be seen as a succession of one-parameter class interactions with local unitary corrections to account for permutation of the parameters e.g. $e^{-i(\alpha_x\sigma_x\otimes\sigma_x+\alpha_z\sigma_z\otimes\sigma_z)}\equiv e^{-i\alpha_z\sigma_z\otimes\sigma_z}(H\otimes H)e^{-i\alpha_x\sigma_x\otimes\sigma_x}(H\otimes H)\equiv C\text{-}R_{\hat{z}}(4\alpha_z)(H\otimes H)C\text{-}R_{\hat{z}}(4\alpha_x)$ (up to local unitary corrections). Viewing it this way demonstrates why it should not have a strictly deterministic set of parameters but also why the form of the unitary actions brought about is always $R_{\hat{z}}(\beta)R_{\hat{x}}(\gamma)=H J(\beta)J(\gamma) H$. Considering this the issue for using the two parameter class is just the inability to make a deterministic choice between two pairs of $\{\beta,\gamma\}$ to produce a universal single qubit gate set. It is similar in effect to flipping the resource requirements for the one-parameter case in time so that rather than preparing in the $\{|j\rangle\}$ basis, we only measure the ancilla in that basis. The resulting gates on the register would still be the same $\{U_0,U_1\}$ as described before but randomly generated with probabilities $p_0, p_1$ dependent on the initial ancilla state.

In fact, we can generate an approximation to any arbitrary single qubit unitary gate in this way. The stochastically generated sequence $\prod_k U_{i(k)}$ will perform a random walk on the compact set of unitaries and be guaranteed by Poincar\'{e} recurrence to reach an approximation of any unitary eventually. The generation of the strings $\prod_k U_{i(k)}$, though probabilistic, are still consistent with the conditions of the Solovay-Kitaev theorem~\cite{NielsenChuang} and so the efficiency of these approximations might be improved.

Future research will look at finding the gate count required to hit a target averaged over all unitaries for a given error and the scaling with the error. In addition to the relevance to our work, it may serve as a benchmark for any circuit compilation sequence: to see how much better inputting it through a string of ancilla is compared with than ``measuring them all and letting God sort it out''.

\begin{figure}[!h]
\centering
\subfloat[Subfigure 1 list of figures text][ $H_R \Delta(0,0,\frac{\pi}{16})$]{
	\includegraphics[width=0.5\textwidth]{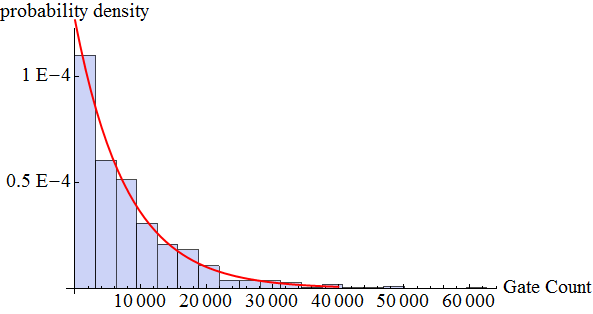}
	\label{fig:OneParameterLowRes}
}
\subfloat[Subfigure 1 list of figures text][ $ \Delta(\frac{\pi}{16},0,\frac{\pi}{16})$]{
	\includegraphics[width=0.5\textwidth]{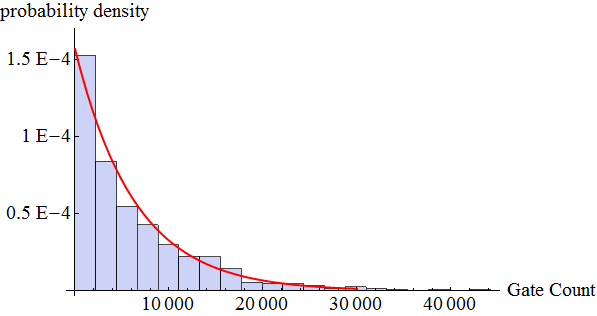}
	\label{fig:TwoParameterLowRes}
}
\caption{Probability distribution of required gate count to achieve target $U_T=R_{\hat{x}}(\frac{\pi}{2})$. a) Use of a single parameter interaction in a 20 bin histogram, b) Use of a two parameter interaction in a 20 bin histogram. The probability distribution corresponding to the exponential distribution parametrised by the mean of the results is displayed by the solid red curve (colour online).}
\label{fig:GateCountsOfStochasticADQC}
\end{figure}

As a preliminary investigation, we simulated the random generation of gates using the interactions $(H\otimes H) \Delta(0,0,\frac{\pi}{16})$ and $\Delta(\frac{\pi}{16},0,\frac{\pi}{16})$, ancilla preparation state $|+\rangle$ and measurement in the computational state basis. For the one-parameter class interaction the resulting unitary gates are $U_0= H R_{\hat{z}}(\frac{\pi}{8})$,$U_1= H R_{\hat{z}}(-\frac{\pi}{8})$, $p_0=p_1=\frac{1}{2}$; for the two-parameter class,$U_0=  R_{\hat{z}}(\frac{\pi}{8}) R_{\hat{x}}(\frac{\pi}{8})$,$U_1=  R_{\hat{z}}(-\frac{\pi}{8}) R_{\hat{x}}(\frac{\pi}{8})$, $p_0=p_1=\frac{1}{2}$. The target unitary was $U_T=R_{\hat{x}}(\frac{\pi}{2})$. At each step a gate corresponding to the $\{U_0,U_1\}$ of each interaction was multiplied to the product of the previous step starting with the identity operator. The normalised trace distance of the product,$V= \prod_k U_{i(k)}$, and the target unitary was evaluated at each step until within a chosen error size $\epsilon<0.05$.
\begin{equation}
\|V-U_T\|=\sqrt{\frac{2-|\text{Tr}[V^{\dagger}U_T]|}{2}}\leq \epsilon
\end{equation}
The number of gates required for this to occur was collected 1000 times and used to create a probability distribution for the gate count required to reach the target unitary (see figure \ref{fig:GateCountsOfStochasticADQC}).

This small investigation reveals some behaviours to be considered in further research. While the target is not achievable in a single step and there is no finite probability for success per step, the aggregate behaviour over many steps, taken over a large number of simulations, can be modelled as a geometric or discretised exponential distribution. This is true for both one-parameter and two-parameter interactions. Particular target unitaries may cause anomalous effects; the two-parameter case is able to produce an exact solution of the target unitary in 4 steps which causes a large peak in the distribution and then suppresses the probability of a result for several steps after (see figure \ref{fig:TwoParameterHighRes} in Appendix \ref{StochasticADQCGraphs}) but with a large enough bin size the exponential model dominates. This provides some features to test when extending the average over a large number of target unitaries. An average over many random target unitaries may smooth out such effects and verify the general applicability of the exponential distribution model. Under this model, we can calculate the probability of reaching any target unitary within a fixed number of steps and increase the number of steps until it provides a desired fidelity.

\section{Entangling gates}

Given the ability to implement any single qubit unitary gate, universal quantum computation requires that we are also able to implement an entangling two-qubit unitary gate between register qubits. Note that direct interaction between register qubits not allowed, the interaction must be mediated by an ancilla qubit in the ADQC model. ADQC is capable of implementing an entangling gate by the use of one ancilla and two implementations of the same fundamental interaction $E_{AR}$ as used in the implementation of single qubit gates. The measurement of the ancilla results in two outcomes but they are equivalent up to local unitary corrections.
\begin{center}
\includegraphics[height=4.6cm,width=14.85cm]{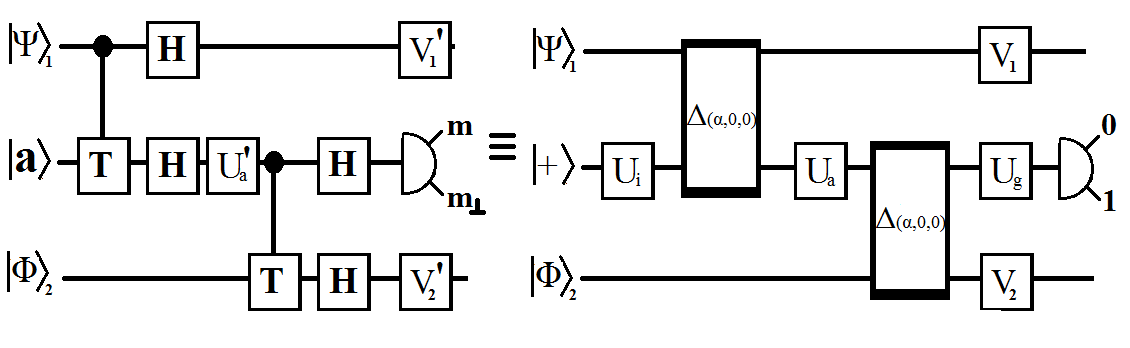}
\captionof{figure}{ The circuit for implementing a two qubit gate and its generalisation. $V_1\otimes V_2$ allows us to make any local unitary corrections to the register qubits. The preparation of the ancilla state is represented by $U_i$ while the choice of measurement basis is represented by $U_g$.}
\label{fig:TwoQubitCircuit}
\end{center}
In our proposal, we attempt the same use of a single ancilla and two interactions but with weaker coupling strength $\Delta(0<\alpha<\frac{\pi}{4})$. We have the freedom to prepare and measure the ancilla in any state/basis, as well as perform local unitary operations and post corrections on it and the register qubits. Due to this, we will focus on the  $e^{-i\alpha\sigma_z\otimes \sigma_z}$ form of the interaction, with any local unitary corrections being incorporated into a single gate; the local H post-interaction unitary on the ancilla can be removed by the appropriate choice of $U_a$---see figure \ref{fig:TwoQubitCircuit} .

\subsection{Interpreting the unitary and entangling conditions}

The operation on the register qubits after measuring the ancilla must be unitary and entangling. This will restrict the unitary operations, prepared states and measurement bases we can use on the ancilla. We need to be able to write the unitary and entangling conditions as some parameter restraints on the circuit. Because the interaction is of the class $\Delta(\alpha,0,0)$, we can make some simplifications. If a register qubit is in a computational basis state $|i\rangle$, $i=0,1$, then it will, through the interaction $\Delta(\alpha,0,0)$ cause a unitary $R_{\hat{z}}((-1)^i 2\alpha)$ action on the ancilla. So for two register qubits there are four potential final ancilla states corresponding to the computational basis $|a_{00}\rangle$ for $|00\rangle_{12}$ etc.. Therefore the transformation of a general register qubit and ancilla state, $|a\rangle |\Phi\rangle_{12}$ will be transformed by the circuit thus:
\begin{equation}
|a\rangle |\Phi\rangle_{12}=|a\rangle\sum_{ij} C_{ij}|ij\rangle_{12} \rightarrow \sum_{ij} C_{ij} |a_{ij}\rangle|ij\rangle_{12}
\end{equation}
Each final ancilla state is a unitary evolution of the original determined by the parameters of the circuit:
\begin{equation}
|a_{ij}\rangle=U_g R_{\hat{z}}((-1)^j 2\alpha) U_a R_{\hat{z}}((-1)^i 2\alpha) |a\rangle
\end{equation}
After a measurement of the ancilla in a state $|m\rangle$ the (unnormalised) state of the register will be
\begin{equation}
\sum_{ij} \langle m|a_{ij}\rangle C_{ij}|ij\rangle_{12}
\end{equation}
The evolution of the register can be represented by a Kraus operator in a matrix representation:
\begin{equation}
\textbf{K}_m=\text{diag}(\langle m|a_{00}\rangle,\langle m|a_{01}\rangle,\langle m|a_{10}\rangle,\langle m|a_{11}\rangle)
\end{equation}

For this operator to be proportional to unitary $|\langle m|_{ij}\rangle|$ must be the same $\forall i,j$. This encapsulates an expression of the equivalence between the unitary condition and the requirement that the measurement of the ancilla extracts no information about the register system. $\langle m|a_{ij}\rangle$=$|\langle m|a_{ij}\rangle|e^{i\phi_{ij}}$; the probability of the measurement does not distinguish between the register states and the term $|\langle m|a_{ij}\rangle|$ drops out and the effective unitary on the register pair of qubits is:
\begin{equation}\label{eq:RegisterTwoQubitUnitary}
\textbf{U}=\text{diag}(
e^{i\phi_{00}}, e^{i\phi_{01}} , e^{i\phi_{10}}, e^{i\phi_{11}})
\end{equation}

The result is also of the class $\Delta(\alpha,0,0)$ and is thus also equivalent to a Control-unitary gate. Being diagonal in the computation basis, we can apply local unitary corrections to convert it into a gate of the form $C-R_{\hat{z}}(\Phi)$. Given a two qubit unitary of the form $ \text{diag}(e^{i \phi_{00}},e^{i \phi_{01}},e^{i \phi_{10}},e^{i \phi_{11}})$ we can multiply it by local unitary gates
\begin{equation}
\left( \begin{array}{cccc}
e^{-i a_1} &\\
& e^{-i a_2} \\
\end{array} \right) \otimes
\left( \begin{array}{cccc}
e^{-i b_1} &\\
& e^{-i b_2} \\
\end{array} \right) = 
\text{diag}(e^{-i(a_1+b_1) }, e^{-i (a_1+b_2)}, e^{-i (a_2+b_1)} ,  e^{-i (a_2+b_2)})
\end{equation}
to give $\text{diag}(e^{i \phi_{00}-(a_1+b_1)}, e^{i \phi_{01}-(a_1+b_2)}, e^{i \phi_{10}-(a_2+b_1)}, e^{i \phi_{11}-(a_2+b_2)})$

We can choose $a_1, b_1, a_2, b_2$ such that
\begin{align}
\phi_{00}-(a_1+b_1)&= 0 \\
\phi_{01}-(a_1+b_2)&= 0 \\
\phi_{10}-(a_2+b_1)&= 0 \\
\rightarrow a_2+b_2&= a_2+b_1-(a_1+b_1)+(a_1+b_2)=\phi_{10}-\phi_{00}+\phi_{01} 
\end{align}

the resulting gate \eqref{eq:RegisterTwoQubitUnitary} must therefore be equivalent to the form
\begin{equation}
\textbf{\~{U}}=\text{diag}(1,1,1,e^{i ((\phi_{11}-\phi_{10})-(\phi_{01}-\phi_{00}})
\end{equation}
We therefore will use $\Phi=\delta\phi_{1}-\delta\phi_{0}=(\phi_{11}-\phi_{10})-(\phi_{01}-\phi_{00})$ to characterise the entangling power of each gate.

\subsection{A geometric picture of the unitary condition}

We have four final ancilla states $|a_{ij}\rangle$ corresponding to the basis states of the register. To allow a measurement basis $\{|m\rangle, |m_{\perp}\rangle\}$ for which $|\langle m|a_{ij}\rangle|$ is constant $\forall i,j$, these four points must all lie in the same plane and form a ring around a cap with $|m\rangle$ at the midpoint. For four $|a_{ij}\rangle$ given by the preparation of the ancilla state, the two ancilla-register couplings and the intermediate local unitary gate on the ancilla, $|m\rangle$ will be fixed and unique. The relative phases of $\langle m|a_{ij}\rangle=|\langle m|a_{ij}\rangle|e^{i\phi_{ij}}$ corresponds to the angles between the points on the ring (see figure \ref{fig:BasicFourPoints}).
\begin{center}
\includegraphics[height=4.5cm,width=8.75cm]{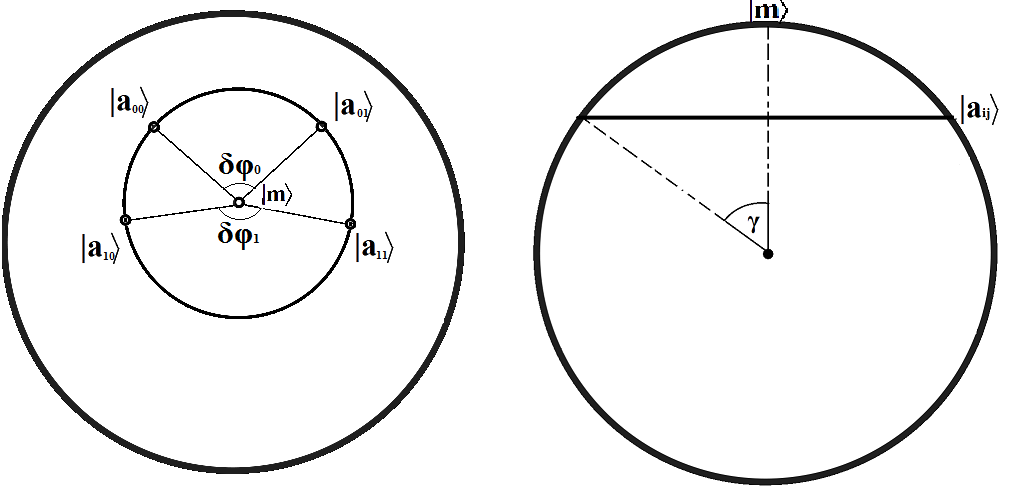}
\captionof{figure}{For four states that have the same value $|\langle m|a_{ij}\rangle|$, there are four points on the Bloch sphere that define a ring that encircle and thus define the state $|m\rangle$. $|\langle m|a_{ij}\rangle|^2=\text{cos}^2(\frac{\gamma}{2})$.}
\label{fig:BasicFourPoints}
\end{center}
\subsubsection{A geometric picture of the entanglement condition}
After the first interaction but before the second the ancilla will be in one of two states $|a_i\rangle$ corresponding to $|\Psi\rangle_1=|i\rangle_1$, $i=0,1$. The second interaction induces a unitary on the second register qubit that is given by $\langle m|\Delta|a_i\rangle$. For entanglement between the first and second register qubit, this unitary must be distinguishable by $i$. Therefore, each $|a_i\rangle$ must have a distinct value of $\langle \sigma_z\rangle_i$. In the geometric picture, two points given by $a_i$ that are on the same horizontal plane, before the second interaction, will produce four $a_{ij}$ on the same horizontal plane afterwards. However, the angle between states marked by different $j$ in the pairs $|a_{0j}\rangle$ and $|a_{1j}\rangle$ would be the same for each $i$. To be entangling $\delta\phi_{1}-\delta\phi_{0}$ must be non-zero.
\begin{center}
\includegraphics[height=4.5cm,width=8.75cm]{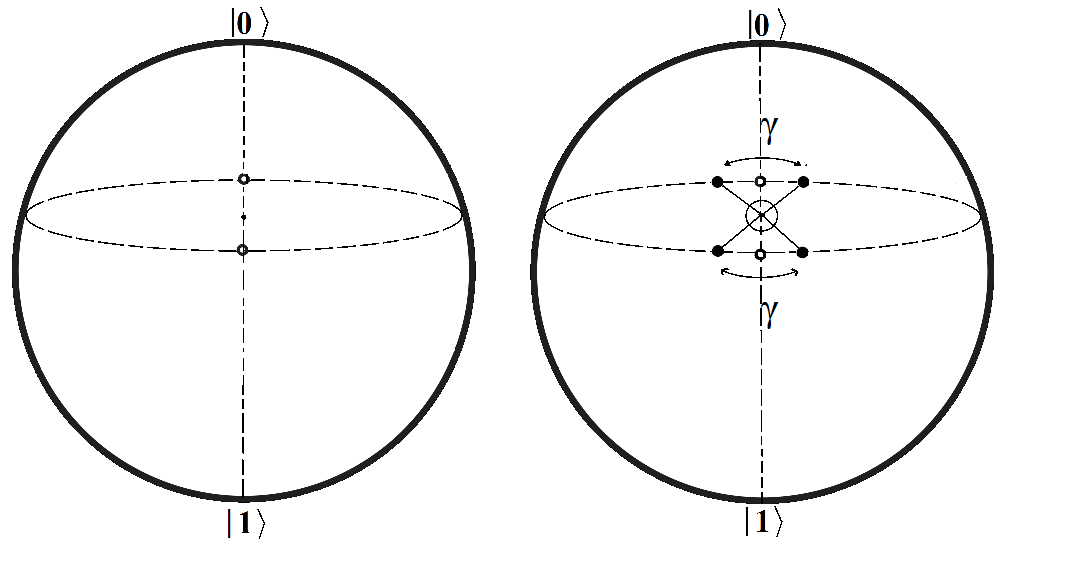}
\captionof{figure}{ All horizontal planes have a cap with a midpoint at the poles thus $|m\rangle= |0\rangle$. An ancilla state in the computational basis corresponds to the same constant unitary enacted with each interaction and extracts no information from the first interaction to transmit in the intermediate stage.}
\label{fig:GeometricPictureOfNonEntanglingUnitary}
\end{center}

From this geometric picture, we can show that the unitary and entangling conditions are fulfilled when the intermediate ancilla states $|a_i\rangle$ are restricted to the same vertical plane (see Appendix \ref{GeometricProofOfUnitaryCondition} and figure \ref{fig:VerticalViewOfFourPoints}).

\subsection{Construction of the ancilla states}
With the intermediate states $|a_i\rangle$ being of the same vertical plane but with some angle between them that we will label as $2\beta$, the final  states $|a_{ij}\rangle$ will be 
\begin{equation}
\text{cos}\left(\frac{\theta-(-1)^i 2\beta}{2}\right)|0\rangle + e^{i(-1)^j 2\alpha}\text{sin}\left(\frac{\theta-(-1)^i 2\beta}{2}\right)
\end{equation}
We can construct the vertically split states $|a_i\rangle$ and manipulate the angle $2\beta$ by the choice of preparation state and intermediate unitary $U_a$. Rotations around the $z$ axis will form a gauge transformation since $|m\rangle$ is specified by $\{|a_{ij}\rangle\}$ so we can, without a loss of generality, assume that the ancilla preparation state is on the $x\text{--}z$ plane: $|a\rangle=\text{cos}(\frac{\theta^\prime}{2})|0\rangle +\text{sin}(\frac{\theta^\prime}{2})|1\rangle$. After the first interaction this becomes $|a\rangle=\text{cos}(\frac{\theta^\prime}{2})|0\rangle +e^{i(-1)^i 2\alpha}\text{sin}(\frac{\theta^\prime}{2})|1\rangle$. Now the solid angle between the two resulting points is not given by $2\alpha$  but by $2\beta$ where $\text{sin}(2\beta)=\text{sin}(\theta^\prime)\text{sin}(2\alpha)$. There will always exist a unitary that can rotate the two points such that they lie in the $x\text{--}z$ plane: $|a\rangle=\text{cos}(\frac{\theta^\prime}{2})|0\rangle +e^{i(-1)^i 2\alpha}\text{sin}(\frac{\theta^\prime}{2})|1\rangle \rightarrow \text{cos}(\frac{\theta-(-1)^i 2\beta}{2})|0\rangle +\text{sin}(\frac{\theta-(-1)^i 2\beta}{2})$. In the geometric picture, this is a rotation around the point where the great circle that connects the two points and the $x\text{--}z$ plane cross, followed by any rotation around $\hat{y}$ of our choice so that $\theta$ is a parameter under control (see figure \ref{fig:ConstructionOfFourPoints}). The next interaction produces the points $\vec{a}_{ij}$ which are, under this order of construction, dependent on the intermediate unitary and the ancilla preparation state.

\begin{center}
\includegraphics[height=4cm,width=14.9cm]{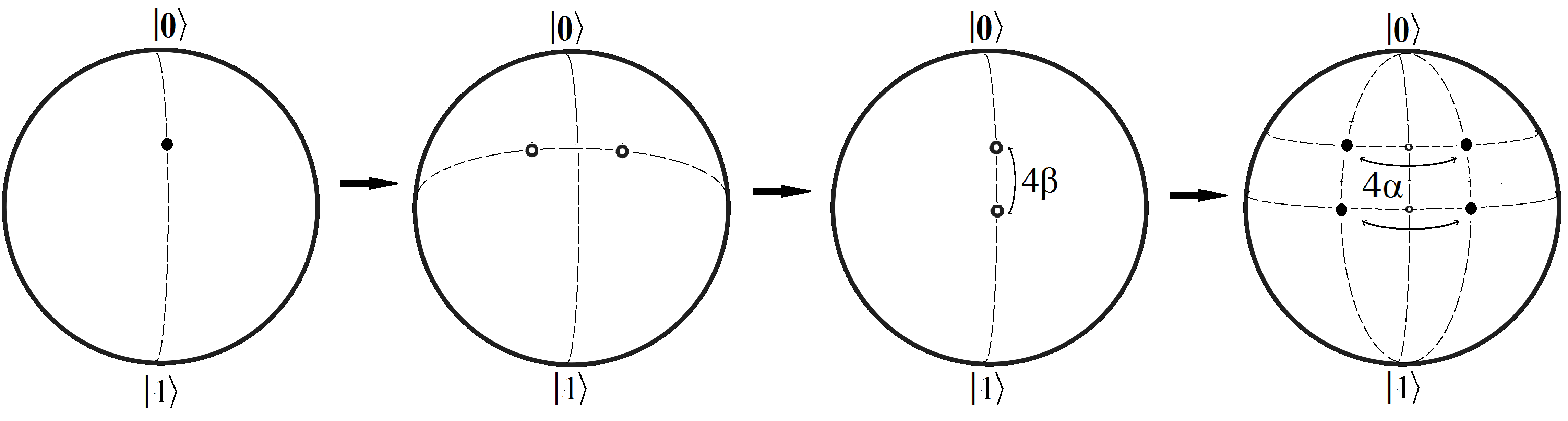}
\captionof{figure}{A geometric representation of the method of constructing an entangling two qubit unitary. The first step corresponds to the ancilla preparation, the second and fourth to the interactions with register qubits and the third to unitary actions on the ancilla in between the interactions.}
\label{fig:ConstructionOfFourPoints}
\end{center}

\subsection{ An example of the relative entanglement powers of the Kraus operators}
As an example, take the intermediate state to have
$\theta=\frac{\pi}{2}$ so that the $|a_i\rangle$ are vertically split
about the $|+\rangle$ state. It is helpful to think of the
intermediate state as $R_{\hat{x}}(\frac{\pi}{2})R_{\hat{z}}(\pm
2\beta)|+\rangle$ while the ancilla was prepared in $|+\rangle$. The
effect of the preparation choice and the reduced solid angle are
treated like an effective reduction in the interaction strength of the
first interaction while the intermediate
$U_a=R_{\hat{x}}(\frac{\pi}{4})$. The four states
\begin{equation}
|a_{ij}\rangle=R_{\hat{z}}((-1)^j 2\alpha)
R_{\hat{x}}(\frac{\pi}{2})R_{\hat{z}}((-1)^i 2\beta)|+\rangle
\end{equation}
will be all symmetrically placed around $|+\rangle$ so we can also say
$|m\rangle=|+\rangle$ and measure in the $\{|+\rangle, |-\rangle\}$
basis.

\begin{equation}
 \langle m|a_{ij}\rangle = \langle +|R_{\hat{z}}((-1)^j 2\alpha)R_{\hat{x}}\left(\frac{\pi}{2}\right)R_{\hat{z}}((-1)^i 2\beta)|+\rangle
\end{equation}
\begin{align}
\textbf{R}_{\hat{z}}(2\alpha)\textbf{R}_{\hat{x}}\left(\frac{\pi}{2}\right)\textbf{R}_{\hat{z}}(2\beta) &= \frac{1}{\sqrt{2}}\left(
\begin{array}{cc}
e^{-iA} & -ie^{-iB} \\
-ie^{iB} & e^{iA} 
\end{array} \right) \\
\rightarrow \langle +|a_{00}\rangle & =\frac{1}{2\sqrt{2}}(e^{-iA} -ie^{-iB} -ie^{iB} + e^{iA})=\frac{1}{\sqrt{2}}(\text{cos}(A)-i\text{cos}(B)) \\
\langle -|a_{00}\rangle & =\frac{1}{2\sqrt{2}}(e^{-iA} -ie^{-iB} +ie^{iB} - e^{iA})=\frac{1}{\sqrt{2}}(-i\text{sin}(A)-\text{sin}(B))
\end{align}
\begin{wraptable}{r}{68mm}
\begin{tabular}{|c|c|c|c|c|c|} \hline
i & j & A$\rightarrow$ & B$\rightarrow$ &$\phi_{ij}^+$&$\phi_{ij}^-$ \\ \hline
0 & 0 & A & B &$\phi_{00}^+$ & $\phi_{00}^-$\\ 
0 & 1 &-B &-A &$-\phi_{00}^+-\frac{\pi}{2}$ & $-\phi_{00}^--\frac{\pi}{2}$ \\ 
1& 0 &B & A & $-\phi_{00}^+-\frac{\pi}{2}$ & $-\phi_{00}^-+\frac{\pi}{2}$ \\ 
1 & 1 &-A &-B &$\phi_{00}^+$ & $\phi_{00}^-$\\ \hline
\end{tabular}
\caption{Table of transformations of A and B for different computational states of the register}
\end{wraptable}
where we define $A=\alpha+\beta$, $B=\alpha-\beta$. Each other element of the two qubit evolution operator will just be a transformation of $A$ and $B$ (which thus fulfils the unitary condition) and because of the $\pm$ symmetries of sine and cosine we will be able to express the final Kraus operator and $\Phi$ in terms of just $\phi_{00}$.
\begin{align}
\Phi_+&=\delta\phi_1-\delta\phi_0= 4\phi_{00}^+ +\pi \\
\Phi_-&= 4\phi_{00}^-+\pi
\end{align}
\begin{center}
\includegraphics[height=4.84cm,width=7.48cm]{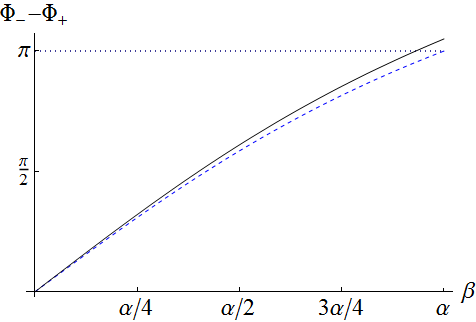}
\captionof{figure}{By manipulating the effective coupling strength, $\beta$, with the ancilla state preparation, the difference in $\Phi$ for the two possible operator outputs can be adjusted.For $\alpha =\frac{\pi}{16}$, the difference between outputs is plotted with the solid (black) curve while the value of $\Phi_-$ is plotted with the dashed (blue) curve (colour online). At $\beta\approx 0.183$, $\delta\Phi=\pi$}
\label{fig:DifferenceInControlPhaseGate}
\end{center}

\subsection{Repeat-until-success entangling gate generation (EGG)}
Without any optimisation of the ancilla initial state and measurement, the generation of entangling gates may, at the very least, be able to perform a random walk (a classical random walk, not to be confused with quantum walks) through the group of Control-$R_{\hat{z}}(\gamma)$. With optimisation, we can apply a protocol to convert the output into a ``success/fail'' scenario with failure corresponding to generation of identity.

In this protocol EGG is enacted twice: once as described before and once again with local unitary changes that convert $\Phi \rightarrow-\Phi$. The latter can be done either by local Pauli-X gate pre- and post-corrections on the register or, if there is access to the ancilla between interactions, by making a correction to the intermediate ancilla unitary to exchange the resulting intermediate states $|a_0\rangle$---$|a_1\rangle$. The change in the second implementation enacts a change of the resulting gate $C\text{-}R_{\hat{z}}(\gamma)\rightarrow C\text{-}R_{\hat{z}}(-\gamma)$. If these two output gates are (under local operator corrections) $C\text{-}R_{\hat{z}}(\Phi_0)$ and $C\text{-}R_{\hat{z}}(\Phi_1)$  with probabilities $p_0$ and $p_1$ then after two two EGG the Control-Unitary is one of $C\text{-}R_{\hat{z}}(\Phi_0-\Phi_1)$, $C\text{-}R_{\hat{z}}(\Phi_1-\Phi_0)$ or $\mathbb{I}$. If $\Phi_0-\Phi_1=\pi$ then we have performed probabilistic CZ generation with a success probability of  $2p_0p_1$.

\begin{center}
\includegraphics[height=5.43cm,width=12cm]{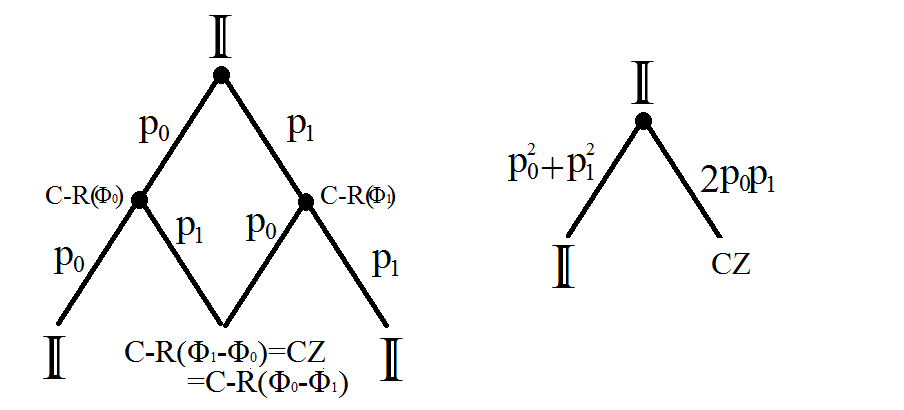}
\captionof{figure}{Since all enacted gates are of the same group, Control-$R_{\hat{z}}(\Phi)$ gates,  multiple gates can be easily combined. Random gate production is a Markovian process. By manipulating the relative values, the output can be limited to a finite probability tree, including a case equivalent to a single outcome ``success/fail'' gate suitable for a repeat-until-success method.}
\label{fig:ProbabilityTreeOfEntanglingGate}
\end{center}

For example, in figure \ref{fig:DifferenceInControlPhaseGate}, we see that we can lower $x$ to match such conditions. In figure \ref{fig:ProbabilityOfSuccessFailGate}, the value of $2p_0p_1$ is calculated, giving us a value of $p\approx 0.128$.

\begin{center}
\includegraphics[height=4.84cm,width=7.48cm]{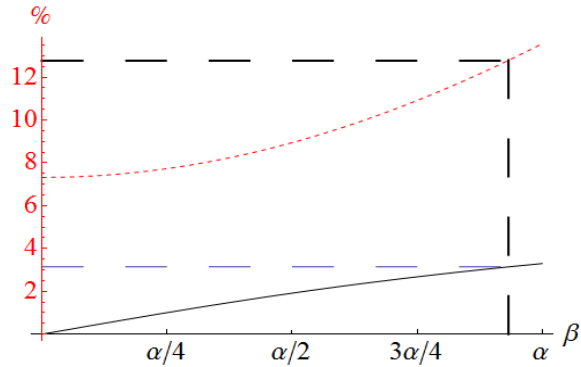}
\captionof{figure}{The probability of achieving two opposing ancilla measurements, $2p_0p_1$, against $\beta$, displayed against the resulting $\Phi$. At the case where $\delta\Phi=\pi$, $p\approx 0.128$ (colour online).}
\label{fig:ProbabilityOfSuccessFailGate}
\end{center}

We provide this one adaptation as a single example of the general principle of using a multi-step protocol to adapt the behaviour of a Markov chain. Though an arbitrary interaction strength leads to the probabilistic generation of gates where the outputs are not local unitary correctable, they are of the same interaction class and thus can be adapted to the group of Control-Phase Rotations and generate group members. Specific physical parameter constraints and choice of target gates may lead to methods that involve different schemes and different groups of gates such as $C\text{-}R_{\hat{z}}(\frac{m\pi}{n})$, $m,n\in\mathbb{Z}$ for a specific $n$ or a continuous parameter group $C\text{-}R_{\hat{z}}(\gamma)$. We provide a method that allows one to enact a repeat-until success method similar to other probabilistic gate proposals for linear optical systems~\cite{RepeatUntilSuccessLinearOptics,ClusterStateQCUsingProbabilisticGates,LinearOpticsQCReview}. 

\section{Initialisation-Measurement}

In ADQC, the ability to perform a projective measurement with an ancilla qubit was straightforward and generalized measurements could be performed by introducing a second ancilla system. With a weak coupling, enacting a projective measurement is a less straightforward proposition requiring many steps and possible fidelity loss. We will provide a protocol for doing so in a naive adaptation of the Control-Z+Hadamard example, followed by discussion of potential future investigation and improvement. 

With a general register qubit $|\Psi\rangle_R=\alpha |0\rangle + \beta|1\rangle$ and ancilla in initialised state $|+\rangle_A$, the interaction $E$ acting on $|\Psi\rangle_R|+\rangle_A$ would produce $\alpha|+\rangle_R|0\rangle_A+\beta|-\rangle_R|1\rangle_A$. A z basis measurement on the ancilla would replicate a z basis measurement on the register state before the interaction; this can also be used to initialise the register qubit with a result $|j\rangle$ producing a register qubit in the state $Z^j|+\rangle$. Local unitary gate preparations on the register could be used to enable measurements or state initialisation in other abses. 

In our proposal, the interaction between register and ancilla can be weaker, enabling us to perform only non-projective measurements. Fortunately Oreskhov \& Brun~\cite{WeakMeasurementsAreUniversal} show that weak measurements are universal and provides an explicit construction for how to use weak measurements to achieve projective and general measurements in a random walk, for two outcome measurements. This can be extended to higher dimensions~\cite{WeakMeasurementsAreUniversal} and it has been shown that one can build up many result general measurements from two output measurements~\cite{BinarySearchTrees}. Thus we expect that it should be possible to achieve at least an approximation of projective measurements using multiple successive non-projective measurements from a single ancilla. We have a specific interest in being able to do so using our interaction of interest $(H_A\otimes H_R). C\text{-}T_{AR}$.

\subsection{Initialisation and Measurement in logarithmic time with a fixed interaction}
At the heart of our proposal is the capability to perform a two qubit unitary that is equivalent up to local unitary gates to a Control-$R_z(\theta)$ gate. Consider $|\Psi\rangle=\alpha|0\rangle+\beta|1\rangle$ and an ancilla in the $|+\rangle$ state:

\begin{equation}
(H_A\otimes H_R).C\text{-}R_z(\theta)|\Psi\rangle_R|+\rangle_A=H (\alpha|0\rangle+\beta \text{cos}\left(\frac{\theta}{2}\right) |1\rangle)_R|0\rangle_A +\beta\text{sin}\left(\frac{\theta}{2}\right)H|1\rangle_R|0\rangle_A
\end{equation}

Now if we measure the ancilla qubit in the $\{|0\rangle,|1\rangle\}$ eigenstate basis and make a local Hadamard correction on the register, the register qubit will be projected into the respective states $\frac{1}{\sqrt{N}}(\alpha|0\rangle+\beta\text{cos}(\frac{\theta}{2})|1\rangle)$ (with probability $N=|\alpha|^2+|\beta|^2 \text{cos}^2(\frac{\theta}{2})$) and $|1\rangle$ (with probability $|\beta|^2\text{sin}^2(\frac{\theta}{2})$).

Consider the effect of many such measurements in the case where every successive result on the ancilla is $|0\rangle$. The iterated effect over $n$ steps is for the register qubit to be projected into the state $\frac{1}{\sqrt{N_n}}(\alpha|0\rangle+\beta\text{cos}^n(\frac{\theta}{2})|1\rangle)$ (where $N_n$ is the appropriate normalization factor). As $n$ increases the state tends to $|0\rangle$ exponentially. The probability of achieving such a chain is $N_n=|\alpha|^2+|\beta|^2 \text{cos}^{2n}(\frac{\theta}{2})$ which will tend to $|\alpha|^2$. Thus we can replicate a projective measurement in $n$ steps up to a state fidelity error of $\beta \text{cos}^n(\frac{\theta}{2})\leq  \text{cos}^n(\frac{\theta}{2})$ with the following procedure:

Set a desired state fidelity error $\epsilon$. Calculate $n \in \mathbb{Z}$ such that $n \geq ln(\epsilon)/ln(\text{cos}(\frac{\theta}{2}))$. Prepare ancilla qubit in the $|+\rangle$ state, couple to register qubit with the interaction $E_{AR}$, measurement ancilla qubit in computational basis, apply $H$ gate correction to register qubit. Repeat until measurement has been performed $n$ times or until ancilla is measured as $|1\rangle$. Label a result of $|0_1 0_2 0_3...0_n\rangle_A$ as $|0\rangle_R$, any other result where the process was terminated by a $|1\rangle_A$ as $|1\rangle_R$. 
\begin{center}
\includegraphics[height=3.2cm,width=10.4cm]{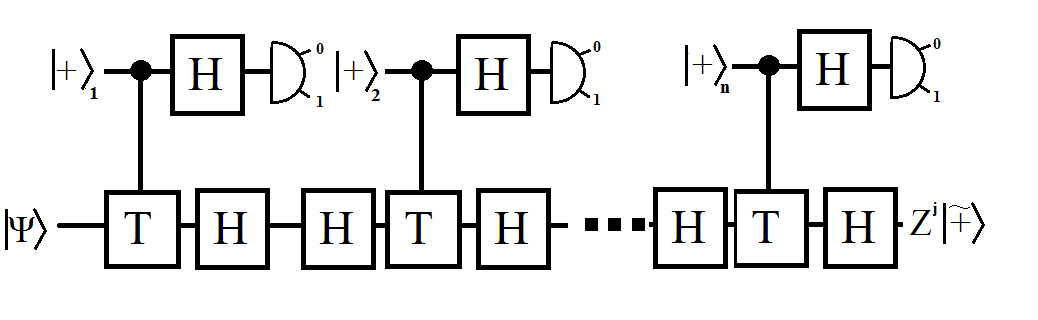}
\captionof{figure}{The circuit description of the iterative measurement protocol using the fundamental interaction $E_{AR}=(H_{a}\otimes H_{R})C\text{-}T$. The local unitary gate components of the interaction require a Hadamard gate correction; though not depicted, this is implemented using ancilla qubits as described in the previous section, leading to a total number of ancilla qubits and interactions of $2n$.}
\label{fig:InitialisationMeasurementCircuit}
\end{center}
In this way, the measurement does not follow a random walk but occurs within a finite bound on the number of steps that is logarithmic with respect to the state fidelity error.

\section{Avenues of further investigation}
Both the classes $\Delta(\alpha,0,0)$ and $\Delta(\alpha_x,\alpha_y,0)$ can be used to enact proportional-to-unitary Kraus operators however our work has mainly dealt with the former. Stochastic ADQC motivates extending the method of enacting two qubit gates to the two parameter interaction class by providing means in which the $\Delta(\alpha_x,\alpha_y,0)$  may be viable for single qubit gates. It also may provide a useful tool for examining the trade off between control or time. The random gate generation method we have described incoporates little control over the initial ancilla state or measurement basis. Potentially future work could consider employing feedback to optimise the choice of ancilla state preparation and measurement basis in each step, resulting in a variable step size and guided random walk. 
The measurement we perform in each step is biased and we incorporate no feedback in our protocol save for the decision to continue or halt based on the previous measurement result. We have also only considered measurement on one qubit while the use of an ancilla provides us with a natural system for making higher dimensional measurements and generalised measurements. Combes \& Jacobs~\cite{CombesJacobs} considered using feedback to speed up the purification of states via continuous measurements. A case where the measurement basis is kept unbiased to the state density operator through feedback is faster than measurement alone~\cite{CombesEtAl}. Much work has been done considering the case where we have a sequence of weak measurement with feedback enacted through unitary operations between steps in time for various goals and measures~\cite{ReconsideringRapidQubitPurification,WisemanBoutenOptimality}. Jacobs~\cite{JacobsProjectQubitFaster} raised the question of whether there is a trade off between speed of projection and loss of initial information. We propose the work of coming up with an unbiased basis feedback protocol for our specific interaction and non-continuous discrete protocol for future investigation.

\section{Conclusion and summary}
We have show how, by relaxing the requirement for stepwise determinism
up to Pauli corrections, a broader class of interactions can be used
to implement a form of ancilla driven quantum computation.  This is
achieved by gate approximation, repeat-until-success strategies, and
generalised measurement.  Single qubit unitary gates can still be
implemented deterministically while two qubit gates and measurement
and state initialisation will be probabilistic and require the
development of probabilistic protocols which we provide examples
of. We provide a measurement process that is logarithmic to the bound
on state fidelity error and reduces to a deterministic approximation
of projective measurements, with the potential for further speed up to
be examined in future research. The conditions for the generation of a
two qubit entangling unitary gate are described in geometric terms. We
demonstrate how choice of ancilla state preparation and local
unitaries enable us to manipulate the entangling strength of the
output gates. This allows us to convert the result into a
``repeat-until-success'' gate. Further research may reveal other
protocols based around adaptations of the entangling gate
power. Additionally, probabilistic protocols for enacting
specific unitary gates may also aid in expanding possible interactions
to include the class $\Delta(\alpha_x,\alpha_y,0)$.

We would like to acknowledge discussions with Dan Browne, Erika Andersson and Shojun Nakayama, and comments on the draft manuscript by Janet Anders and Elham Kashefi. DKLO and KHS are supported by the Quantum Information Scotland (QUISCO) network.

\bibliographystyle{unsrt}
\bibliography{bibliography}

\begin{thebibliography}{10}

\bibitem{GBQC}
D.~Deutsch.
\newblock Quantum computational networks.
\newblock {\em Proceedings of the Royal Society of London. A. Mathematical and
  Physical Sciences}, 425(1868):73--90, 1989.

\bibitem{RaussendorfBriegel}
Robert Raussendorf and Hans~J. Briegel.
\newblock A one-way quantum computer.
\newblock {\em Phys. Rev. Lett.}, 86:5188--5191, May 2001.

\bibitem{QuantumComputationByAdiabaticEvolution}
Edward Farhi, Jeffrey Goldstone, Sam Gutmann, and Michael Sipser.
\newblock {Quantum Computation By Adiabatic Evolution}.
\newblock {\em eprint arXiv:quant-ph/0001106}, January 2000.

\bibitem{KitaevAnyons}
A.Yu. Kitaev.
\newblock Fault-tolerant quantum computation by anyons.
\newblock {\em Annals of Physics}, 303(1):2 -- 30, 2003.

\bibitem{ADQC}
Janet Anders, Daniel K.~L. Oi, Elham Kashefi, Dan~E. Browne, and Erika
  Andersson.
\newblock Ancilla-driven universal quantum computation.
\newblock {\em Phys. Rev. A}, 82(2):020301, Aug 2010.

\bibitem{AluminiumOpticalClock}
C.~W. Chou, D.~B. Hume, J.~C.~J. Koelemeij, D.~J. Wineland, and T.~Rosenband.
\newblock Frequency comparison of two high-accuracy ${\mathrm{al}}^{+}$ optical
  clocks.
\newblock {\em Phys. Rev. Lett.}, 104:070802, Feb 2010.

\bibitem{Ohshima2000}
Toshio Ohshima.
\newblock All-optical electron spin quantum computer with ancilla bits for
  operations in each coupled-dot cell.
\newblock {\em Phys. Rev. A}, 62:062316, Nov 2000.

\bibitem{QuantumRegisterNVCentre}
M.~V.~Gurudev Dutt, L.~Childress, L.~Jiang, E.~Togan, J.~Maze, F.~Jelezko,
  A.~S. Zibrov, P.~R. Hemmer, and M.~D. Lukin.
\newblock Quantum register based on individual electronic and nuclear spin
  qubits in diamond.
\newblock {\em Science}, 316(5829):1312--1316, 2007.

\bibitem{UlmNVCentreAncilla}
A.~Bermudez, F.~Jelezko, M.~B. Plenio, and A.~Retzker.
\newblock Electron-mediated nuclear-spin interactions between distant
  nitrogen-vacancy centers.
\newblock {\em Phys. Rev. Lett.}, 107:150503, Oct 2011.

\bibitem{BlinovEtAl04}
B.~B. Blinov, D.~L. Moehring, L.-M. Duan, and C.~Monroe.
\newblock Observation of entanglement between a single trapped atom and a
  single photon.
\newblock {\em Nature}, 428:153--157, Mar 2004.

\bibitem{FlyingQubits}
G.~Cordourier-Maruri, F.~Ciccarello, Y.~Omar, M.~Zarcone, R.~de~Coss, and
  S.~Bose.
\newblock Implementing quantum gates through scattering between a static and a
  flying qubit.
\newblock {\em Phys. Rev. A}, 82:052313, Nov 2010.

\bibitem{TwistedGraphStatesADQC}
E.~Kashefi, D.K.L. Oi, D.~Browne, J.~Anders, and E.~Andersson.
\newblock Twisted graph states for ancilla-driven universal quantum
  computation.
\newblock {\em Electronic Notes in Theoretical Computer Science}, 249(0):307 --
  331, 2009.
\newblock Proceedings of the 25th Conference on Mathematical Foundations of
  Programming Semantics (MFPS 2009).

\bibitem{BrownBriegelOWC}
Dan Browne and Hans~J. Briegel.
\newblock One-way quantum computation.
\newblock In {\em Lectures on Quantum Information}, pages 359--358. Wiley-VCH,
  Weinheim,Germany, 2007.

\bibitem{NovelSchemesForMBQC}
D.~Gross and J.~Eisert.
\newblock Novel schemes for measurement-based quantum computation.
\newblock {\em Phys. Rev. Lett.}, 98:220503, May 2007.

\bibitem{MBQCBeyondOW}
D.~Gross, J.~Eisert, N.~Schuch, and D.~Perez-Garcia.
\newblock Measurement-based quantum computation beyond the one-way model.
\newblock {\em Phys. Rev. A}, 76:052315, Nov 2007.

\bibitem{OiSchirmer2012}
D.~K.~L. Oi and S.~G. Schirmer.
\newblock Quantum system characterization with limited resources.
\newblock {\em Philosophical Transactions of the Royal Society A: Mathematical,
  Physical and Engineering Sciences}, 370(1979):5386--5395, 2012.

\bibitem{ADQCWithTwisted}
J.~Anders, E.~Andersson, D.E. Browne, E.~Kashefi, and D.K.L. Oi.
\newblock Ancilla-driven quantum computation with twisted graph states.
\newblock {\em Theoretical Computer Science}, 430(0):51 -- 72, 2012.
\newblock Mathematical Foundations of Programming Semantics (MFPS XXV).

\bibitem{NielsenChuang}
Michael~A. Nielsen and Isaac~L. Chuang.
\newblock {\em Quantum Computation and Quantum Information}.
\newblock Cambridge University Press, Cambridge, United Kingdom, 2000.

\bibitem{KeylWernerCM}
M.~Keyl and R.F. Werner.
\newblock Channels and maps.
\newblock In {\em Lectures on Quantum Information}, pages 73--86. Wiley-VCH,
  Weinheim,Germany, 2007.

\bibitem{StinespringCSA}
W.~Forrest Stinespring.
\newblock Positive functions on c*-algebras.
\newblock {\em Proceedings of the American Mathematical Society}, 6(2):pp.
  211--216, 1955.

\bibitem{KrausCirac01}
B.~Kraus and J.~I. Cirac.
\newblock Optimal creation of entanglement using a two-qubit gate.
\newblock {\em Phys. Rev. A}, 63:062309, May 2001.

\bibitem{GeometricTheoryOfNonLocal}
Jun Zhang, Jiri Vala, Shankar Sastry, and K.~Birgitta Whaley.
\newblock Geometric theory of nonlocal two-qubit operations.
\newblock {\em Phys. Rev. A}, 67:042313, Apr 2003.

\bibitem{Rezakhani}
A.~T. Rezakhani.
\newblock Characterization of two-qubit perfect entanglers.
\newblock {\em Phys. Rev. A}, 70:052313, Nov 2004.

\bibitem{OnUniversalAndFaultTolerant}
P.~Oscar Boykin, Tal Mor, Matthew Pulver, Vwani Roychowdhury, and Farrokh
  Vatan.
\newblock {On universal and fault-tolerant quantum computing}.
\newblock {\em eprint arXiv:quant-ph/9906054}, June 1999.

\bibitem{AliferisGottesmanPreskill}
Panos Aliferis, Daniel Gottesman, and John Preskill.
\newblock {Quantum accuracy threshold for concatenated distance-3 codes}.
\newblock {\em eprint arXiv:quant-ph/0504218}, April 2005.

\bibitem{RepeatUntilSuccessLinearOptics}
Yuan~Liang Lim, Almut Beige, and Leong~Chuan Kwek.
\newblock Repeat-until-success linear optics distributed quantum computing.
\newblock {\em Phys. Rev. Lett.}, 95:030505, Jul 2005.

\bibitem{ClusterStateQCUsingProbabilisticGates}
D.~Gross, K.~Kieling, and J.~Eisert.
\newblock Potential and limits to cluster-state quantum computing using
  probabilistic gates.
\newblock {\em Phys. Rev. A}, 74:042343, Oct 2006.

\bibitem{LinearOpticsQCReview}
Pieter Kok, W.~J. Munro, Kae Nemoto, T.~C. Ralph, Jonathan~P. Dowling, and
  G.~J. Milburn.
\newblock Linear optical quantum computing with photonic qubits.
\newblock {\em Rev. Mod. Phys.}, 79:135--174, Jan 2007.

\bibitem{WeakMeasurementsAreUniversal}
Ognyan Oreshkov and Todd~A. Brun.
\newblock Weak measurements are universal.
\newblock {\em Phys. Rev. Lett.}, 95:110409, Sep 2005.

\bibitem{BinarySearchTrees}
Erika Andersson and Daniel K.~L. Oi.
\newblock Binary search trees for generalized measurements.
\newblock {\em Phys. Rev. A}, 77:052104, May 2008.

\bibitem{CombesJacobs}
Joshua Combes and Kurt Jacobs.
\newblock Rapid state reduction of quantum systems using feedback control.
\newblock {\em Phys. Rev. Lett.}, 96:010504, Jan 2006.

\bibitem{CombesEtAl}
Joshua Combes, Howard~M. Wiseman, Kurt Jacobs, and Anthony~J. O'Connor.
\newblock Rapid purification of quantum systems by measuring in a
  feedback-controlled unbiased basis.
\newblock {\em Phys. Rev. A}, 82:022307, Aug 2010.

\bibitem{ReconsideringRapidQubitPurification}
H~M Wiseman and J~F Ralph.
\newblock Reconsidering rapid qubit purification by feedback.
\newblock {\em New Journal of Physics}, 8(6):90, 2006.

\bibitem{WisemanBoutenOptimality}
Howard~M. Wiseman and Luc Bouten.
\newblock Optimality of feedback control strategies for qubit purification.
\newblock {\em Quantum Information Processing}, 7:71--83, 2008.

\bibitem{JacobsProjectQubitFaster}
Kurt Jacobs.
\newblock How to project qubits faster using quantum feedback.
\newblock {\em Phys. Rev. A}, 67:030301, Mar 2003.

\end{thebibliography}

\appendix
\appendixpage
\section{Restrictions of the parameters of the ancilla found in a geometric proof} \label{GeometricProofOfUnitaryCondition}
Each point can be represented by the spherical coordinates that describe the state $\vec{a}_k=(\theta_k,\phi_k)$, $k=1,2,3,4$. For out notation the values of $k$ will correspond to the values expressed by $ij$ in binary.

Each point can be expressed in Cartesian coordinates by the relationships:
\begin{align}
|\vec{a}_k.\vec{x}|&=\text{sin}(\theta_k)\text{cos}(\phi_k) \\
|\vec{a}_k.\vec{y}|&=\text{sin}(\theta_k)\text{sin}(\phi_k) \\
|\vec{a}_k.\vec{z}|&=\text{cos}(\theta_k)
\end{align}

Three points alone can always be found to be on the same plane. We will define a plane from three points and then find the expression for the distance from the fourth point to that plane. Thus we will find the conditions for the fourth point to be in the same plane as the other three. The equation for a plane defined by three points is
\begin{equation}
a.x+b.y+c.z+d=0
\end{equation}
\begin{equation}
a=\frac{-d}{D}\left|
\begin{array}{ccc}
1 &y_1 &z_1 \\
1 &y_2 &z_2 \\
1 &y_3 &z_3 
\end{array}\right| ,\\
b=\frac{-d}{D}\left|
\begin{array}{ccc}
x_1 &1 &z_1 \\
x_2 &1 &z_2 \\
x_3 &1 &z_3 
\end{array}\right| ,\\
c=\frac{-d}{D}\left|
\begin{array}{ccc}
x_1 &y_1 &1 \\
x_2 &y_2 &1 \\
x_3 &y_3 &1 
\end{array}\right| ,\\
D=\left|
\begin{array}{ccc}
x_1 &y_1 &z_1 \\
x_2 &y_2 &z_2\\
x_3 &y_3 &z_3 
\end{array}\right|
\end{equation}
There is a freedom of choice of $d$ so it can be set $d=D$. The distance of point 4 to the plane is given by
\begin{equation}
\text{distance}=\frac{|a.x_4+b.y_4+c.z_4+d|}{\sqrt{a^2+b^2+c^2}}
\end{equation}
As we are only interested in the case where the distance is zero, we can ignore the normalisation factor and simply examine
\begin{equation}
\text{distance}^{\prime}=|a.x_4+b.y_4+c.z_4+d|
\end{equation}

We now make use of some restrictions on the formation of these four points. Before the second interaction, there are intermediate ancilla states corresponding to the computational basis of the first register qubit. The final four points are constructed from these two points by the second interaction which imparts a rotation around the $\vec{z}$ axis by $\pm 2\alpha$ with $\pm$ corresponding to the second qubit being in the computational basis state $|j\rangle$, $j=0,1$. This means that we can set 
\begin{equation}\label{eq:FixedHorizontalPairs}
\theta_1=\theta_2, \theta_3=\theta_4
\end{equation}
and 
\begin{equation} \label{eq:FixedAzimuthalDifference}
\phi_2-\phi_1=\phi_4-\phi_3=4\alpha
\end{equation}

Using \eqref{eq:FixedHorizontalPairs}, the distance between the fourth point and the plane can be simplified to

\begin{equation}
2(\text{cos}(\theta_2)-\text{cos}(\theta_4))\left[\text{cos}\left(\phi_2-\frac{\phi_3+\phi_4}{2}\right)-\text{cos}\left(\phi_1-\frac{\phi_3+\phi_4}{2}\right)\right]\text{sin}(\theta_2)\text{sin}(\theta_4)\text{sin}(\frac{\phi_3-\phi_4}{2})
\end{equation}

This generates several possible conditions for the fourth point to lie in the plane, some more trivial than others. If $\text{cos}(\theta_2)=\text{cos}(\theta_4)$ then all four points must lie on the same horizontal plane which means that there is no entangling power. sin$(\theta_2)=0$ and sin$(\theta_4)=0$ would mean that there are only three distinct points with one being at the pole i.e. the $|0\rangle$ state. Due to the construction of these four points it is not possible for $\phi_3-\phi_4=0$ to be true. The final condition is that 
\begin{equation}
\text{cos}\left(\phi_2-\frac{\phi_3+\phi_4}{2}\right)=\text{cos}\left(\phi_1-\frac{\phi_3+\phi_4}{2}\right)
\end{equation}
If \eqref{eq:FixedAzimuthalDifference} is then substituted in, this becomes
\begin{align}
\text{cos}\left(\phi_1+4\alpha -\frac{2\phi_3+4\alpha}{2}\right)&=\text{cos}\left(\phi_1 -\frac{2\phi_3+4\alpha}{2}\right) \\
\text{cos}\left(\phi_1-\phi_3+2\alpha\right)&=\text{cos}\left(\phi_1 -\phi_3-2\alpha\right) 
\end{align}
Since $\alpha$ is non-zero, this requires $\phi_1=\phi_3+n\pi$ for $n\in \mathbb{Z}$ i.e. the two points are in the same vertical plane.

\begin{center}
\includegraphics[height=4.27cm,width=4.51cm]{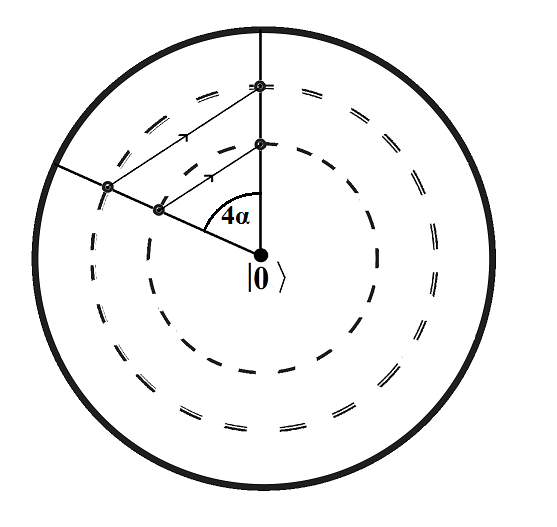}
\captionof{figure}{A 2d projection of the construction of four point on the Bloch sphere. By restricting the points to only one of two polar and one of two azimuthal angles, the vectors that connect two points of the same polar angle will be parallel. This guarantees that all four points lie on the same plane.}
\label{fig:VerticalViewOfFourPoints}
\end{center}

\newpage

\section{Graphs of stochastic ADQC gate count and the exponential distribution} \label{StochasticADQCGraphs}

\begin{figure}[!h]
\centering
\subfloat[Subfigure 1 list of figures text][ $H_R \Delta(0,0,\frac{\pi}{16})$]{
	\includegraphics[width=0.5\textwidth]{OneParameter11_LowResHistogram.png}
	\label{fig:OneParameterLowRes}
}
\subfloat[Subfigure 1 list of figures text][ $ \Delta(\frac{\pi}{16},0,\frac{\pi}{16})$]{
	\includegraphics[width=0.5\textwidth]{TwoParameter11_LowResHistogram.png}
	\label{fig:TwoParameterLowRes}
}

\subfloat[Subfigure 1 list of figures text][ $H_R \Delta(0,0,\frac{\pi}{16})$]{
	\includegraphics[width=0.5\textwidth]{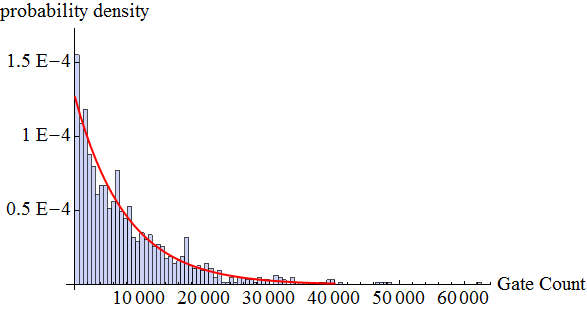}
	\label{fig:OneParameterHighRes}
}
\subfloat[Subfigure 1 list of figures text][ $ \Delta(\frac{\pi}{16},0,\frac{\pi}{16})$]{
	\includegraphics[width=0.5\textwidth]{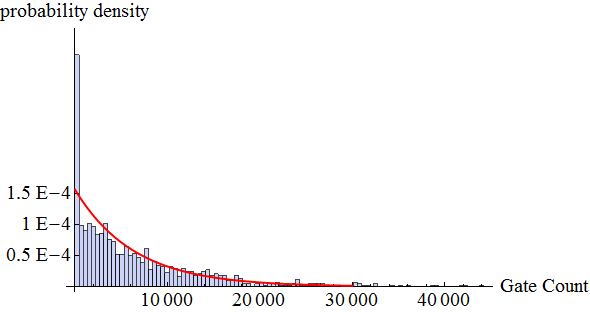}
	\label{fig:TwoParameterHighRes}
}
\caption{Probability distribution of required gate count to achieve target $U_T=R_{\hat{x}}(\frac{\pi}{2})$. a) Use of a single parameter interaction in a 20 bin histogram, b) Use of a two parameter interaction in a 20 bin histogram,c) Use of a single parameter interaction in a 100 bin histogram, d) Use of a two parameter interaction in a 100 bin histogram; note the large aberration in the first division. The probability distribution corresponding to the exponential distribution parametrised by the mean of the results is displayed by the solid red curve (colour online).}
\label{fig:GateCountsOfStochasticADQCWithHighResolution}
\end{figure}
\begin{figure}[!h]
\centering
\subfloat[Subfigure 1 list of figures text][ $H_R \Delta(0,0,\frac{\pi}{16})$]{
	\includegraphics[width=0.5\textwidth]{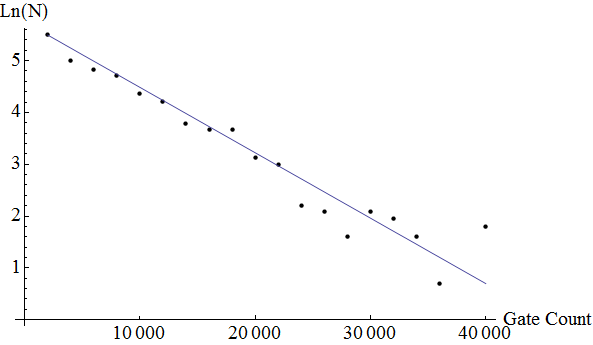}
	\label{fig:OneParameterLogOfBin}
}
\subfloat[Subfigure 1 list of figures text][ $ \Delta(\frac{\pi}{16},0,\frac{\pi}{16})$]{
	\includegraphics[width=0.5\textwidth]{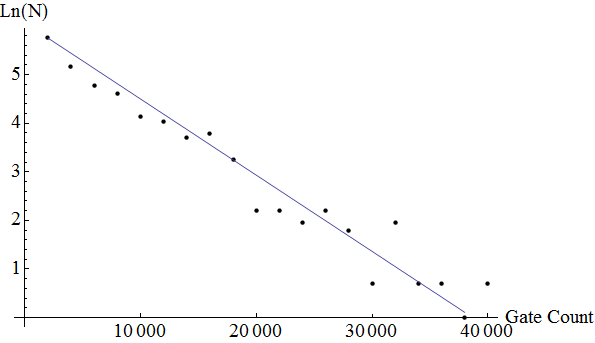}
	\label{fig:TwoParameterLogOfBin}
}
\caption{The natural logarithm of the bin counts of the number of gates required to achieve  target $U_T=R_{\hat{x}}(\frac{\pi}{2})$. By taking the natural logarithm the exponential distribution model forms a linear curve which fits the points generated by the simulation. Noise distrupts the linear behaviour for very low counts of high gate number. Figure a matches the one parameter interaction, figure b matches the two parameter interaction.}
\label{fig:LogOfBinVsGate}
\end{figure}
\end{document}